\documentclass[twocolumn,groupedaddress,showpacs,nofootinbib]{revtex4}
\usepackage{graphicx}
\usepackage{dcolumn}% Align table columns on decimal point
\usepackage{bm}% bold math
\usepackage{amssymb}
\usepackage{mathrsfs}
\usepackage{amsmath}
\usepackage{epsfig}
\usepackage{here}
\usepackage[dvips]{color}
\usepackage{hhline}

\begin{document}

\title{ATIC and PAMELA Results on Cosmic $e^\pm$ Excesses and Neutrino Masses}

\author{Xiao-Jun Bi$^{1}_{}$}
\email{bixj@ihep.ac.cn}

\author{Pei-Hong Gu$^{2}_{}$}
\email{pgu@ictp.it}

\author{Tianjun Li$^3$}
\email{tli@itp.ac.cn}

\author{Xinmin Zhang$^{4}_{}$}
\email{xmzhang@ihep.ac.cn}

\affiliation{$^{1}_{}$Key Laboratory of Particle Astrophysics,
Institute of High Energy Physics, Chinese Academy of Sciences,
Beijing 10049, China \& Center for High Energy Physics,
Peking University, Beijing 100871, P.R. China\\
$^{2}_{}$The Abdus Salam International Centre for Theoretical
Physics, Strada Costiera 11, 34014 Trieste, Italy
\\
$^3$ Institute of Theoretical Physics, Chinese Academy of
Sciences, Beijing 100080, P. R. China, and George P. and Cynthia
W. Mitchell Institute for Fundamental Physics, Texas A$\&$M
University, College Station, TX 77843, USA\\
 $^{4}_{}$Theoretical Physics Division, Institute of High
Energy Physics, Chinese Academy of Sciences, Beijing 100049, P.R.
China}

\begin{abstract}

Recently the ATIC and PAMELA collaborations released their results
which show the abundant $e^{\pm}_{}$ excess in cosmic rays well
above the background, but not for the $\bar{p}$. Their data if
interpreted as the dark matter particles' annihilation imply that
the new physics with the dark matter is closely related to the
lepton sector. In this paper we study the possible connection of the
new physics responsible for the cosmic $e^{\pm}_{}$ excesses to the
neutrino mass generation. We consider a class of models and do the
detailed numerical calculations. We find that some models can
account for the ATIC and PAMELA $e^{\pm}_{}$ and $\bar{p}$ data and
at the same time generate the small neutrino masses.

\end{abstract}

\pacs{14.60.Pq, 95.35.+d  \hfill [MIFP-08-35]}

\maketitle

\subsection{Introduction}

Recently the ATIC \cite{Chang:2008zz} and PPB-BETS
\cite{Torii:2008} collaborations have reported the
electron/positron spectrum measurement up to $\sim 1$~TeV, with an
obvious bump above the background at $\sim 300-800\,\textrm{GeV}$
and $\sim 500-800\,\textrm{GeV}$, respectively. At the same time,
the PAMELA collaboration also released their first cosmic-ray
measurements on the positron fraction \cite{Adriani:2008zr} and
the $\bar{p}/p$ ratio \cite{Adriani:2008zq}. The positron fraction
shows the significant excesses above $10\,\textrm{GeV}$ up to
$\sim 100\,\textrm{GeV}$, compared with the background predicted
by the conventional cosmic-rays propagation model. This result is
consistent with the previous measurements by HEAT
\cite{Barwick:1997ig} and AMS \cite{Aguilar:2007yf}. Clearly, the
ATIC, PPB-BETS and PAMELA results indicate the existence of a new
source of primary electrons and positrons while the hadronic
processes are suppressed. It is well known that the annihilation
of the dark matter (DM) can be a possible origin for the primary
cosmic rays, which can account for the ATIC/PPB-BETS and PAMELA
data simultaneously \cite{yin}. However, the $\bar{p}/p$ ratio
gives strong constraint on the nature of dark matter particles.
The widely discussed lightest neutralino in supersymmetric models
as dark matter candidate may be difficult to explain the PAMELA
data due to the constraints \cite{yin,donato}. These constraints
may imply some special relations between the DM sector and
leptons.

Another interesting topic related to the leptons is the neutrino
mass-generation. In order to understand the small but nonzero
neutrino masses, required by various neutrino oscillation
experiments, we can consider the elegant seesaw \cite{minkowski1977}
extension of the standard model (SM). The original seesaw scenario
is realized by integrating out the heavy particles at tree level
\cite{minkowski1977,mw1980}. As an alternative, one may consider the
radiative seesaw models \cite{ma1998,ma2006,gs2007,gs2008} where the
small neutrino masses are generated at loop level.

In this paper we connect the leptonic annihilation of the DM to
the mass generation of the neutrinos. We find that in some
radiative seesaw models the neutral Majorana or Dirac fermions as
the DM can dominantly annihilate into the charged leptons and
neutrinos. For proper choice of model parameters and boost factor,
the leptonic annihilations of the DM can explain the
electron/positron excesses observed by the ATIC/PPB-BETS and
PAMELA cosmic-ray experiments. With small modifications, we might
explain the INTEGRAL~\cite{Churazov:2004as} experiment via the
exciting DM~\cite{Finkbeiner:2007kk}. In addition, we propose one
kind of models with traditional seesaw
mechanism~\cite{minkowski1977}. These models can explain the
ATIC/PPB-BETS and PAMELA cosmic-ray experiments similarly, and
explain the INTEGRAL experiment as well.

\subsection{Effective Lagrangian}

\begin{figure}
\vspace{4cm} \epsfig{file=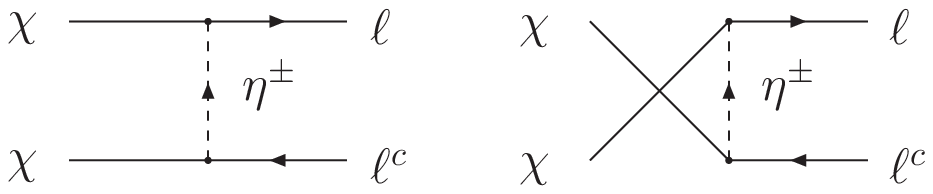, bbllx=3cm,
bblly=6.0cm, bburx=13cm, bbury=16cm, width=7cm, height=7cm, angle=0,
clip=0} \vspace{-8.5cm} \caption{\label{annihilation-majorana} The
annihilations of the Majorana $\chi$ into the charged leptons
through the exchange of $\eta^{\pm}_{}$.}
\end{figure}

\begin{figure}
\vspace{9cm} \epsfig{file=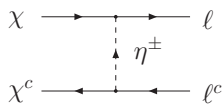, bbllx=-0.3cm,
bblly=6.0cm, bburx=9.7cm, bbury=16cm, width=7cm, height=7cm,
angle=0, clip=0} \vspace{-8.5cm} \caption{\label{annihilation-dirac}
The annihilations of the Dirac $\chi$ into the charged leptons
through the exchange of $\eta^{\pm}_{}$.}
\end{figure}

We start with our discussions with an
effective Lagrangian
\begin{eqnarray}
\label{effective-lagrangian} \mathcal{L}\supset
-y_{\ell}^{}\overline{\ell^{-}_{L,R}}\eta^{-}_{}\chi
+\textrm{H.c.}\,,
\end{eqnarray}
where $\ell=e,\mu,\tau$ are the charged leptons, $\eta^{\pm}_{}$ is
a scalar carrying the same charge as $\ell^{\pm}_{}$, $\chi$
denotes a neutral fermion. We further assume $\eta^{\pm}_{}$ and
$\chi$ to have no other Yukawa couplings with the SM particles.
Therefore, in case of $m_{\chi}^{}<m_{\eta^{\pm}_{}}^{}$, $\chi$ can
dominantly annihilate into the charged leptons since the loop-order
annihilations of $\chi$ into quarks and/or photon are highly
suppressed. Fig. \ref{annihilation-majorana} shows the annihilations
of the Majorana $\chi$ into the charged leptons while Fig.
\ref{annihilation-dirac} is for the Dirac $\chi$.

The cross sections of $\chi$ annihilating into the charged leptons
are given by
\begin{subequations}
\begin{widetext}
\begin{eqnarray}
\label{cross-section-majorana-1} \sigma_{\alpha\beta}^{} v &\equiv&
\sigma_{\chi\chi\rightarrow \ell_{\alpha}^{}\ell^{c}_{\beta}}^{} v
\nonumber\\
&=& \frac{1}{32\pi}|y_{\alpha}^{}|^{2}_{}|y_{\beta}^{}|^{2}_{}
\frac{1}{s\sqrt{s\left(s-4m_{\chi}^{2}\right)}}\left\{\sqrt{s\left(s-4m_{\chi}^{2}\right)}\right.\nonumber\\
&&+
\left[2\left(m_{\eta^{\pm}_{}}^{2}-m_{\chi}^{2}\right)+\frac{2m_{\chi}^{2}s}{s+2m_{\eta^{\pm}_{}}^{2}-2m_{\chi}^{2}}\right]
\ln\left|\frac{s+2m_{\eta^{\pm}_{}}^{2}-2m_{\chi}^{2}-\sqrt{s\left(s-4m_{\chi}^{2}\right)}}{s+2m_{\eta^{\pm}_{}}^{2}-2m_{\chi}^{2}+\sqrt{s\left(s-4m_{\chi}^{2}\right)}}\right|\\
&&\left.+2\left(m_{\eta^{\pm}_{}}^{2}-m_{\chi}^{2}\right)^{2}_{}\left[\frac{1}{s+2m_{\eta^{\pm}_{}}^{2}-2m_{\chi}^{2}-\sqrt{s\left(s-4m_{\chi}^{2}\right)}}
-\frac{1}{s+2m_{\eta^{\pm}_{}}^{2}-2m_{\chi}^{2}+\sqrt{s\left(s-4m_{\chi}^{2}\right)}}\right]\right\}\nonumber\,,
\end{eqnarray}
\begin{eqnarray}
\label{cross-section-dirac-1} \sigma_{\alpha\beta}^{} v &\equiv&
\sigma_{\chi\chi^{c}_{}\rightarrow
\ell_{\alpha}^{}\ell^{c}_{\beta}}^{} v \nonumber\\
&=& \frac{1}{32\pi}|y_{\alpha}^{}|^{2}_{}|y_{\beta}^{}|^{2}_{}
\frac{1}{s\sqrt{s\left(s-4m_{\chi}^{2}\right)}}\left\{\sqrt{s\left(s-4m_{\chi}^{2}\right)}\right.\nonumber\\
&&+ 2\left(m_{\eta^{\pm}_{}}^{2}-m_{\chi}^{2}\right)
\ln\left|\frac{s+2m_{\eta^{\pm}_{}}^{2}-2m_{\chi}^{2}-\sqrt{s\left(s-4m_{\chi}^{2}\right)}}{s+2m_{\eta^{\pm}_{}}^{2}-2m_{\chi}^{2}+\sqrt{s\left(s-4m_{\chi}^{2}\right)}}\right|\\
&&\left.+2\left(m_{\eta^{\pm}_{}}^{2}-m_{\chi}^{2}\right)^{2}_{}\left[\frac{1}{s+2m_{\eta^{\pm}_{}}^{2}-2m_{\chi}^{2}-\sqrt{s\left(s-4m_{\chi}^{2}\right)}}
-\frac{1}{s+2m_{\eta^{\pm}_{}}^{2}-2m_{\chi}^{2}+\sqrt{s\left(s-4m_{\chi}^{2}\right)}}\right]\right\}\nonumber\,,
\end{eqnarray}
\end{widetext}
\end{subequations}
where $\chi$ is a Majorana or Dirac particle, respectively. Here
$v$ is the relative velocity between the two annihilating particles
in their cms system. Up to $\mathcal{O}(v^{2}_{})$, the cross
sections (\ref{cross-section-majorana-1}) and
(\ref{cross-section-dirac-1}) can be respectively simplified as
\begin{subequations}
\begin{widetext}
\begin{eqnarray}
\label{cross-section-majorana-2} \sigma_{\alpha\beta}^{}
v&\simeq&\frac{1}{128\pi}|y_{\alpha}^{}|^{2}_{}|y_{\beta}^{}|^{2}_{}
\frac{1}{(2+r)^{2}_{}}v^{2}_{}\frac{1}{m_{\chi}^{2}} \quad\textrm{in Majorana case}\,,\\
\label{cross-section-dirac-2} \sigma_{\alpha\beta}^{}
v&\simeq&\frac{1}{128\pi}|y_{\alpha}^{}|^{2}_{}|y_{\beta}^{}|^{2}_{}
\left\{\frac{4}{(2+r)^{2}_{}}+\left[\frac{1}{(2+r)^{2}_{}}-\frac{4}{(2+r)^{3}_{}}\right]v^{2}_{}\right\}\frac{1}{m_{\chi}^{2}}
\quad\textrm{in Dirac case}\,,
\end{eqnarray}
\end{widetext}
\end{subequations}
with the definition,
\begin{eqnarray}
r\equiv \frac{m_{\eta^{\pm}_{}}^{2}-m_{\chi}^{2}}{m_{\chi}^{2}}>0\,.
\end{eqnarray}
By inputting $v \sim 10^{-3}_{}$ (the average velocity in our
galaxy) and $r=0$, we further derive
\begin{subequations}
\begin{widetext}
\begin{eqnarray}
\label{cross-section-majorana-3} \langle\sigma_{\alpha\beta}^{}
v\rangle&\lesssim& 1.5\times
10^{-32}_{}\,\textrm{cm}^{3}_{}\textrm{sec}^{-1}_{}\left(\frac{700\,\textrm{GeV}}{m_{\chi}^{}}\right)^{2}_{}|y_{\alpha}^{}|^{2}_{}|y_{\beta}^{}|^{2}_{}
\quad\textrm{in Majorana case}\,,\\
\label{cross-section-dirac-3} \langle\sigma_{\alpha\beta}^{}
v\rangle&\lesssim& 5.9\times
10^{-26}_{}\,\textrm{cm}^{3}_{}\textrm{sec}^{-1}_{}\left(\frac{700\,\textrm{GeV}}{m_{\chi}^{}}\right)^{2}_{}|y_{\alpha}^{}|^{2}_{}|y_{\beta}^{}|^{2}_{}\quad\textrm{in
Dirac case}\,,
\end{eqnarray}
\end{widetext}
\end{subequations}
where $y_{e,\mu,\tau}$ should be smaller than  ${\sqrt {4\pi}}$ for a perturbative theory.

In the following, we will embed the effective Lagrangian
Eq.~(\ref{effective-lagrangian}) into some radiative seesaw models where
the tiny neutrino masses arise from loop contributions. And we
can embed the effective Lagrangian Eq.~(\ref{effective-lagrangian})
into the models with traditional seesaw mechanism as well. For our
convention, we denote  the SM lepton doublets, right-handed leptons
and Higgs doublet as $\psi_{L_\alpha}^{}(\textbf{2},-\frac{1}{2})=
\displaystyle{\left(\nu_{L_\alpha}^{},\ell_{L_\alpha}^{-}\right)^{T}_{}}$,
$e_{R_\alpha}^{}(\textbf{1},-1)$ and
$\phi(\textbf{2},-\frac{1}{2})=\displaystyle{(\phi^{0}_{},\phi^{-}_{})^{T}_{}}$,
respectively, where their $SU(2)_L\times U(1)_Y$ quantum numbers are
given as well.

%%%%%%%%%%%%%%%%%%%%%%%%%%%%%%%%%%%%%%%%%%%%%%%%%%%%%%%%%%%%%%%%%%%%%%

%%%%%%%%%%%%%%%%%%%%%%%%%%%%%%%%%%%%%%%%%%%%%%%%%%%%%%%%%%%%%%%%%%%%%%

\subsection{Model I with Majorana Neutrino Masses and Mixings
from Radiative Corrections}

\begin{figure}
\vspace{7cm} \epsfig{file=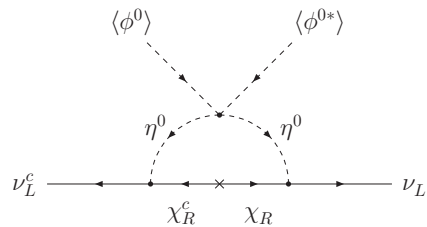, bbllx=5.3cm,
bblly=6.0cm, bburx=15.3cm, bbury=16cm, width=6.5cm, height=6.5cm,
angle=0, clip=0} \vspace{-8.5cm} \caption{\label{mass-generation-1}
The loop diagram for generating small masses of Majorana neutrinos.}
\end{figure}

We consider a Majorana radiative seesaw model~\cite{ma1998, ma2006}
where one introduces three right-handed neutrinos $\chi^{}_{R_i}$
and a SM scalar doublet $\eta$  with $SU(2)_L\times U(1)_Y$
quantum numbers $(\mathbf{2}, -\frac{1}{2})$. In this model,
$\chi^{}_{R_i}$ and $\eta$ are odd under a ${\bf Z}_{2}^{}$
symmetry, while the SM fields carry even parity. Respect to this
${\bf Z}_{2}^{}$, the scalar doublet $\eta$ will not develop nonzero
vacuum expectation value (VEV). Therefore, as shown in Fig.
\ref{mass-generation-1}, the neutrino masses can be radiatively
generated through the following Yukawa couplings, the Majorana mass
term of the right-handed neutrinos, and the quartic interaction of
scalars
\begin{eqnarray}
\label{majorana-seesaw} -\mathcal{L} \supset {1\over 2}
m_{\chi_{i}}^{} \overline{\chi_{R_i}^{c}} \chi^{}_{R_i} + y_{\alpha
i}^{}\overline{ \psi_{L_\alpha}^{} } \eta \chi_{R_i}^{}  +
\lambda(\phi^{\dagger} \eta)^2 + \textrm{H.c.} \,.
\end{eqnarray}
Here we have conveniently chosen the base, in which the Majorana
mass matrix of the right-handed neutrinos is diagonal and real, i.e.
$m_{\chi}^{}=\textrm{Diag}\{m_{\chi_1}^{}, m_{\chi_2}^{},
m_{\chi_3}^{}\}$. Obviously, the above Lagrangian in Eq.
(\ref{majorana-seesaw}) can induce the effective theory in Eq.
(\ref{effective-lagrangian}). If the lightest $\chi$ is lighter than
$\eta_{R}^{}=\frac{1}{\sqrt{2}}\displaystyle{\textrm{Re}(\eta^{0}_{})}$,
$\eta_{I}^{}=\frac{1}{\sqrt{2}}\displaystyle{\textrm{Im}(\eta^{0}_{})}$
and $\eta^{\pm}_{}$, it will have no decay modes and will dominantly
annihilate into the charged leptons and the neutrinos. The total
cross section of the lightest $\chi$ into the charged leptons and
neutrinos is given by
\begin{eqnarray}
\label{cross-section-majorana-4} \langle\sigma v\rangle\lesssim
3.0\times
10^{-32}_{}\,\textrm{cm}^{3}_{}\textrm{sec}^{-1}_{}\left(\frac{700\,\textrm{GeV}}
{m_{\chi}^{}}\right)^{2}_{}\left[(y^{\dagger}_{}y)_{ii}\right]^{2}_{}\,.
\end{eqnarray}
Here $m_{\eta^{0}_{R}}^{2}\simeq m_{\eta^{0}_{I}}^{2}\simeq
m_{\eta^{\pm}_{}}^{2}$ has been adopted since we are interested in
the case where the Yukawa couplings $y_{\alpha i}^{}$ are large
enough for a desired cross section and the quartic coupling
$\lambda$ is small enough to guarantee the smallness of neutrino
masses.

Because the two light right-handed neutrinos $\chi_1$ and $\chi_2$
can be degenerate, we might explain the
INTEGRAL~\cite{Churazov:2004as} experiment via the exciting
DM~\cite{Finkbeiner:2007kk}. To explain the INTEGRAL experiment by
producing $\chi_2$ via the channel $\chi_1 \chi_1 \to \chi_1
\chi_2$ through the tree-level and ladder
diagrams~\cite{Finkbeiner:2007kk}, we introduce a SM singlet $S$
that is ${\bf Z}_2$ even. The discussions are similar to the
following subsection ${\bf D}$ except that we will have additional
terms in the Lagrangian in addition to that in Eq. (\ref{DAMA-I})
\begin{eqnarray}
 -{\cal L}& =& {1\over 2} \mu^2 S^2
+ {1\over {3!}} A_1 S^3 + {1\over {4!}}\lambda' S^4 \nonumber\\ &&
+ {{\lambda_9}\over 2} A_2 S (\eta^{\dagger} \eta) +
{{\lambda_{10}}\over 2} A_3 S (\phi^{\dagger} \phi)
 +\textrm{H.c.}~,~\,
\end{eqnarray}
where $A_i$ are mass dimension-one parameters. In short, this
approach needs some fine-tuning.

To avoid the fine-tuning, we can consider the large discrete symmetry
$\Gamma$ in the dark matter sector, for example $Z_4$ in
Ref.~\cite{Finkbeiner:2007kk}, or as
discussed in subsection ${\bf D}$. Note that the Majorana masses for the
right-handed neutrinos are forbidden by $\Gamma$ since it is not
${\bf Z}_2$. We will need two SM singlet Higgs fields: one
generates the large Majorana masses, while the other generates the
small mass splitting around 1 MeV and has Yukawa coupling about
$0.18$ or larger to the DM and exciting DM fields so that
we can produce $\chi_2$~\cite{Finkbeiner:2007kk}.
Because this kind of
model is similar to the models in the following
subsection ${\bf C}$, we will not study it in detail here.

\subsection{Model II with Dirac Neutrino Masses and Mixings
from Radiative Corrections}

\begin{figure}
\vspace{6.1cm} \epsfig{file=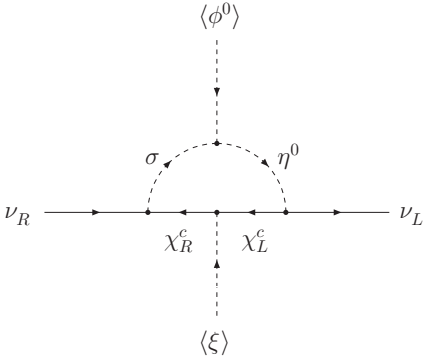, bbllx=5.3cm,
bblly=6.0cm, bburx=15.3cm, bbury=16cm, width=6.5cm, height=6.5cm,
angle=0, clip=0} \vspace{-7.0cm}
\caption{\label{mass-generation-2} The loop diagram for generating
small masses of Dirac neutrinos.}
\end{figure}

Let us consider a Dirac radiative seesaw model~\cite{gs2007} by
extending the SM with a SM scalar doublet $\eta
(\textbf{2},-\frac{1}{2})$, a complex scalar $\xi (\textbf{1},0)$, a
real scalar $\sigma (\textbf{1},0)$, three right-handed neutrinos
$\nu_{R_\alpha}^{} (\textbf{1},0)$ and three Dirac fermions
$\chi_{L_i,R_i}^{} (\textbf{1},0)$. In this model, there is a
$U(1)_{D}^{}$ gauge symmetry under which $\xi$, $\nu_{R_\alpha}^{c}$
and $\chi_{R_i}^{}$ carry the quantum number $\mathbf{1}$. In
addition, $\eta$, $\sigma$ and $\chi_{L_i,R_i}^{}$ are odd while all
of other fields are even under a $Z_2^{}$ symmetry. Moreover,
$\nu_{R_\alpha}^{}$ and $\chi_{L_i,R_i}^{c}$ are arranged for the
same lepton number with the SM leptons. Conserving the
$U(1)_{D}^{}$, $Z_{2}^{}$ and lepton number, we can write down the
following interactions
\begin{eqnarray}
-{\cal L}& \supset& y_{\alpha i}^{}
\overline{\psi_{L_\alpha}^{}}\eta\chi_{L_i}^{c}
+h_{\alpha i}\sigma\overline{\nu_{R_\alpha}^{}}\chi_{R_i}^{c} \nonumber\\
&& + f_{ij}\xi\overline{\chi_{R_i}^{}} \chi_{L_j}^{} + \mu
\sigma\eta^\dagger_{}\phi +\textrm{H.c.}\,. \label{dirac-seesaw}
\end{eqnarray}
Since $\eta^{0}_{}$ and $\sigma$ are forbidden to develop the
nonzero VEVs as a result of the ${\bf Z}_{2}^{}$ protection, the
neutrinos can only obtain their small Dirac masses via the one-loop
diagram, as shown in Fig. \ref{mass-generation-2}. It is
straightforward to see the effective Lagrangian in Eq.
(\ref{effective-lagrangian}) can be embedded into this Dirac
radiative seesaw model. Note there exists a mixing between $\xi$ and
$\phi^0$, which will lead to the co-annihilation of the dark matter
into the quarks. But $\langle\xi\rangle$ should be
about one order larger than
$\langle\phi^0\rangle$ for giving the dark matter a mass close to the TeV
scale. So, the $\xi-\phi^{0}_{}$ mixing, of the order of
$O(\langle\phi^0\rangle/\langle\xi\rangle)\sim 0.1$, could have no
significant implications. For convenience and without loss of
generality, we, in the following, shall choose the basis in which the
Dirac mass matrix of the singlet fermions is diagonal and real, i.e.
$m_{\chi}^{}=f\langle\xi\rangle=\textrm{Diag}\{m_{\chi_1}^{},
m_{\chi_2}^{}, m_{\chi_3}^{}\}$. Similar to that in the Majorana
radiative seesaw model, here the lightest $\chi$ can also dominantly
annihilate into the charged leptons and the neutrinos if it is
lighter than
$\eta_{I}^{}=\frac{1}{\sqrt{2}}\displaystyle{\textrm{Im}(\eta^{0}_{})}$,
$\eta^{\pm}_{}$ and the other two physical states $\rho_{1,2}^{}$
(combinations of
$\eta_{R}^{}=\frac{1}{\sqrt{2}}\displaystyle{\textrm{Re}(\eta^{0}_{})}$
and $\sigma$). The total cross section of this lightest $\chi$ into
the charged leptons and neutrinos is given by
\begin{eqnarray}
\label{cross-section-dirac-4} \langle\sigma v\rangle&\lesssim&
5.9\times
10^{-26}_{}\,\textrm{cm}^{3}_{}\textrm{sec}^{-1}_{}\left(\frac{700\,\textrm{GeV}}
{m_{\chi}^{}}\right)^{2}_{}\nonumber\\
&&\times
\left\{\left[\textrm{Tr}(h^{\dagger}_{}h)_{ii}^{}\right]^{2}_{}+2\left[\textrm{Tr}(y^{\dagger}_{}y)_{ii}^{}\right]^{2}_{}\right\}\,.
\end{eqnarray}
Here $m_{\rho_{1}^{}}^{2}\simeq m_{\rho_{2}^{}}^{2}\simeq
m_{\eta^{0}_{I}}^{2}\simeq m_{\eta^{\pm}_{}}^{2}$ has been adopted
by taking small trilinear coupling $\mu$. This is consistent with
our scope for large Yukawa couplings $y_{\alpha i}$ and $h_{\alpha
i}$ and then a desired cross section.

Naively, we may guess that we do not need to introduce another SM
singlet Higgs field $\xi'$ to explain the INTEGRAL experiment.
However, the Dirac masses for $\chi_{LR}^i$ are generated from the
terms $f_{ij}\xi\overline{\chi_R^{i}} \chi_L^{j}$, and we need two
degenerate Dirac fermions $\chi_{LR}^{\prime i}$ in the mass
basis. Because the masses for dark matter fields is about 600-800
GeV and the mass splitting for the two light Dirac fermions is
about a few MeVs, we obtain that
 the Yukawa coupling constants for
$ \xi \chi_{L}^{\prime 1} \chi_{R}^{\prime 2} $  and $ \xi
\chi_{L}^{\prime 2} \chi_{R}^{\prime 1} $ are about $10^{-5}$ unless
we fine-tune the coefficients $f_{ij}$. Thus, it is difficult to
generate $\chi_{LR}^{\prime 2}$ through the
tree-level and ladder diagrams, and
then explain the INTEGRAL experiment~\cite{Finkbeiner:2007kk}.
Therefore, we introduce another SM singlet Higgs
field $\xi'$ with VEV about 10 MeV. $\xi'$ has $U(1)_D$ charge
$\mathbf{+1}$ and is ${\bf Z}_2$ even. The additional Lagrangian is
\begin{eqnarray}
 -{\cal L}& =&
 {1\over 2} m_{\xi'}^2 \xi^{\prime \dagger} \xi'
+ {{\lambda'_1}\over 4} (\xi^{\prime \dagger} \xi')^2 +
{{\lambda'_2}\over 2} (\xi^{\prime \dagger} \xi')  (\eta^{\dagger}
\eta) \nonumber\\ && + {{\lambda'_3}\over 4}  \sigma^2
(\xi^{\prime \dagger} \xi^{\prime}) + {{\lambda'_4}\over 2}
(\xi^{\prime \dagger} \xi') (\xi^{ \dagger} \xi) +
{{\lambda'_5}\over 2}  (\xi^{\prime \dagger} \xi') (\phi^{\dagger}
\phi) \nonumber\\ && + \left( f'_{ij}\xi' \overline{\chi_R^{i}}
\chi_L^{j}+ \tilde{m}^2 \xi^{\dagger} \xi' + {{\lambda'_6}\over 2}
(\xi^{\dagger} \xi')^2 \right.\nonumber\\ && \left. +
{{\lambda'_7}\over 2} (\xi^{\dagger} \xi') (\xi^{\prime \dagger}
\xi') + {{\lambda'_8}\over 2} (\xi^{\dagger} \xi') (\xi^{ \dagger}
\xi) \right.\nonumber\\ && \left. + {{\lambda'_9}\over 2}
(\xi^{\dagger} \xi') (\eta^{\dagger} \eta) + {{\lambda'_{10}}\over
4} \sigma^2 (\xi^{\dagger} \xi') \right.\nonumber\\ && \left. +
{{\lambda'_{11}}\over 2} (\xi^{\dagger} \xi') (\phi^{\dagger}
\phi)
 +\textrm{H.c.} \right)~.~
\label{AD-dirac-seesaw}
\end{eqnarray}
To explain the INTEGRAL
experiment, we need $f'_{ij}$ to be about $0. 18$ or
larger~\cite{Finkbeiner:2007kk}.

%%%%%%%%%%%%%%%%%%%%%%%%%%%%%%%%%%%%%%%%%%%%%%%%%%%%%%%%%%%%%%%%%%%%%%

%%%%%%%%%%%%%%%%%%%%%%%%%%%%%%%%%%%%%%%%%%%%%%%%%%%%%%%%%%%%%%%%%%%%%%

\subsection{Model  III  with  Neutrino Masses and Mixings from Traditional
 Seesaw Mechanism}

We consider the generalized Standard Model with two or three heavy
right-handed neutrinos. The neutrino masses and mixings are
 generated via the traditional seesaw mechanism, and the
observed baryon asymmetry is explained via leptogensis.

First, we consider the model that can only explain the ATIC and
PAMELA experiments. We introduce dark matter sector that has a SM
singlet scalar field $\widetilde{E}$ with $SU(2)_L\times U(1)_Y$
quantum numbers $(\mathbf{1}, \mathbf{-1})$ and a Majorana fermion
$\chi$.  To have a stable dark matter candidate $\chi$, we
consider the ${\bf Z}_2$ symmetry under which only $\widetilde{E}$ and
$\chi$ are odd. The renormalizable Lagrangian for the dark sector
and its interaction with the SM is
\begin{eqnarray}
\label{NSM-lagrangian} -\mathcal{L} &=& {1\over 2} m_{\chi}
\overline{\chi^c} \chi + {1\over 2} m^2_{E} \widetilde{E}^{\dagger}
\widetilde{E} + {{\lambda_1}\over 4} (\widetilde{E}^{\dagger}
\widetilde{E})^2 \nonumber\\ && + {{\lambda_2}\over 2}
(\widetilde{E}^{\dagger} \widetilde{E}) (\phi^{\dagger} \phi) +
\left(y_e^{i}\overline{ e_R^{i} } \widetilde{E} \chi + \textrm{H.c.}
\right)~,~
\end{eqnarray}
where $m_{\chi}$ and $m_{E}$ are mass terms, and $\lambda_i$ and
$y_{e}^{i}$ are Yukawa couplings. Also, $\lambda_1$ and
$\lambda_2$ are real, while $y_e^{i}$  can be complex and then
violate CP symmetry. To have $\chi$ as a dark matter, we require
that $m_{\chi} < m_{E}$.

Second, we can generalize the dark matter sector in the
 above model to explain the INTEGRAL experiment.
In the dark matter sector, in addition to $\widetilde{E}$ as
above, we introduce a SM singlet scalar field $S$, and two
Majorana particles $\chi_1$ and $\chi_2$ respectively with  left
and right  chiralities~\cite{Finkbeiner:2007kk}. And
the discrere symmetry in the dark
matter sector is ${\bf Z}_4$. Under ${\bf Z}_4$,
the fields in the dark matter
sector transform as follows
\begin{eqnarray}
 \chi_{1,2} \rightarrow e^{\pm i \pi/2} \chi_{1,2} ~,~
\widetilde{E} \rightarrow e^{ -i \pi/2} \widetilde{E}~,~ S
\rightarrow - S ~.~\,
\end{eqnarray}
The renormalizable Lagrangian  is
\begin{eqnarray}
 -{\cal L}& =& m_D \overline{\chi^c_1} \chi_2
+ {1\over 2} m^2_{E} \widetilde{E}^{\dagger} \widetilde{E} +
{1\over 2} m_S^2 S^{\dagger} S + {{\lambda_1}\over 4}
(\widetilde{E}^{\dagger} \widetilde{E})^2 \nonumber\\ && +
{{\lambda_2}\over 4} (S^{\dagger} S)^2 + {{\lambda_3}\over 2}
(S^{\dagger} S) (\widetilde{E}^{\dagger} \widetilde{E})
\nonumber\\ && + {{\lambda_4}\over 2}  (\widetilde{E}^{\dagger}
\widetilde{E})
 (\phi^{\dagger} \phi)
+{{\lambda_5}\over 2} (S^{\dagger} S) (\phi^{\dagger} \phi)
\nonumber\\ && + \left( y_e^{i}\overline{ e_{R}^{i} }
\widetilde{E}^{\dagger} \chi_1 +y_1 S \overline{\chi^c_1} \chi_1 +
y_2 S \overline{\chi^c_2} \chi_2 \right.\nonumber\\ && \left. +
{{\lambda_6}\over 2} S^2 (\widetilde{E}^{\dagger} \widetilde{E}) +
{{\lambda_7}\over 2} S^2 (\phi^{\dagger} \phi)
 +\textrm{H.c.} \right)~.~
\label{DAMA-I}
\end{eqnarray}

In the basis $\chi,\chi_{*} = 1/\sqrt{2}(\chi_1 \mp \chi_2)$, the
mass matrix for $\chi$'s is
\begin{eqnarray}
M=
\begin{pmatrix}
y_+ S - m_D & y_- S \cr y_- S & y_+ S + m_D
\end{pmatrix}~,~
\end{eqnarray}
where $y_\pm = \frac{1}{2}(y_1 \pm y_2)$.  Thus, at the leading
order, the Dirac fermion can be decomposed as two degenerate
Majorana fermions $\chi$ and $\chi^*$. If the ${\bf Z}_4$ symmetry
is broken weakly to ${\bf Z}_2$ by giving  $S$ a VEV, we expect
these states to be split by a small amount $\delta=2 y_+ \langle S
\rangle$. Suppose $m_S^2 < 0$, we obtain that $\langle S^2 \rangle
= - m_S^2/\lambda_2$, and  the mass splitting is then just $\delta
=  y_+ |m_S|/\sqrt{\lambda_2}$~\cite{Finkbeiner:2007kk}. To
explain the INTEGRAL experiment, the mass splitting is about a few
MEV. Note that the couplings $y_{\pm}$ can be as small as around
0.18, and then the VEV of $S$ should be around 10 MeV. Thus, the
mixing between the  Higgs field $\phi^0$ and $S$ is very small
from $\lambda_5$ term, and then the anti-protons from the
$\chi\chi$  annihilation are highly suppressed.

Interestingly, this kind of models can be naturally embedded into
the supersymmetric flipped $SU(5)\times U(1)_X$
models~\cite{Bae:2008sp} and the string or F-theory derived flipped
$SU(5)\times U(1)_X$ models~\cite{Jiang:2008yf}. An important
difference of this kind of models from the other two models is that
it predicts no neutrino signals from DM annihilation. The models I
and II give large neutrino fluxes as well as the positron flux.
Detecting neutrino fluxes from the Galactic Center may discriminate
these models.

%%%%%%%%%%%%%%%%%%%%%%%%%%%%%%%%%%%%%%%%%%%%%%%%%%%%%%%%%%%%%%%%%%%%%%

%%%%%%%%%%%%%%%%%%%%%%%%%%%%%%%%%%%%%%%%%%%%%%%%%%%%%%%%%%%%%%%%%%%%%%

\subsection{Phenomenology}

In this part we study the dark matter models given above in detail
to explain the ATIC and PAMELA data. The propagation of charged
particles in the Galaxy is calculated numerically in order to
compare with data measured at the Earth. The charged particles
propagate diffusively in the Galaxy due to the scattering with
random magnetic field~\cite{Gaisser:1990vg}. The interactions with
interstellar medium  will lead to energy losses of the primary
electrons and positrons. In addition, the overall convection
driven by the Galactic wind and reacceleration due to the
interstellar shock will also affect the distribution and spectrum
of electrons. The propagation equation can be written
as~\cite{Strong:1998pw}
\begin{eqnarray}
\frac{\partial \psi}{\partial t} =Q({\bf
r},p)&+&\nabla\cdot(D_{xx}\nabla \psi-{\bf
V_c}\psi)+\frac{\partial}{\partial p}p^2D_{pp}\frac{\partial}
{\partial p}\frac{1}{p^2}\psi \nonumber \\
 &-& \frac{\partial}{\partial p}\left[\dot{p}\psi
-\frac{p}{3}(\nabla\cdot{\bf V_c}\psi)\right] \ , \label{prop}
\end{eqnarray}
where $\psi$ is the number density of cosmic ray particles per unit
momentum interval, $Q({\bf r},p)$ is the source term, discribing the
primary particles injected into the interstellar meadium, $D_{xx}$
is the spatial diffusion coefficient, ${\bf V_c}$ is the convection
velocity, $D_{pp}$ is the diffusion coefficient in momentum space
used to describe the reacceleration process, $\dot{p}\equiv{\rm
d}p/{\rm d}t$ is the momentum loss rate, mainly induced by
synchrotron radiation and inverse Compon scattering. In the work we
solve Eq. (\ref{prop}) numerically adopting the GALPROP package
\cite{galprop}.

For DM annihilation, the source term of electrons and positrons
have the form
\begin{equation}
\label{anni} Q_{A}({\bf r},E)= BF \frac{<\sigma v>_{A}
\rho^2(r)}{2\,m^2_{DM}} \frac{dN(E)}{dE}\ \ ,
\end{equation}
where, BF is the boost factor, $<\sigma v>_{A}$ is the DM
annihilation cross section, $m_{DM}$ is the mass of DM particle, and
$\frac{dN(E)}{dE}$ is the electron/positron spectrum from one pair
of DM annihilation, $\rho(r)$ is DM  density distribution in the
Galaxy. In this work, we adopt the NFW profile \cite{nfw} for DM
distribution with  the local DM density
$\rho_{\odot}=0.3\,\textrm{GeV}/\textrm{cm}^3$.

\begin{figure}[!htb]
\begin{center}
\scalebox{0.25}{\includegraphics{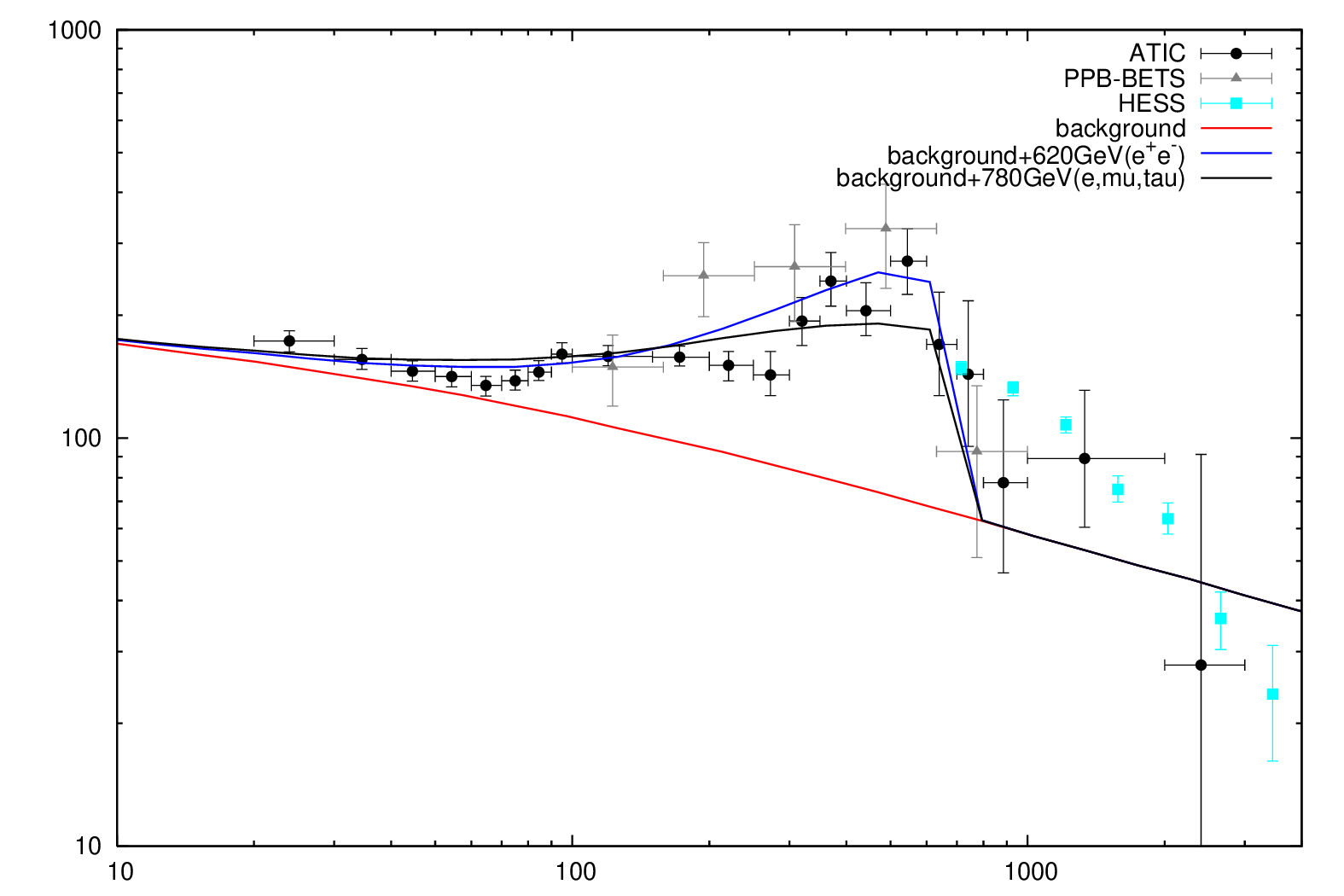}}
\scalebox{0.25}{\includegraphics{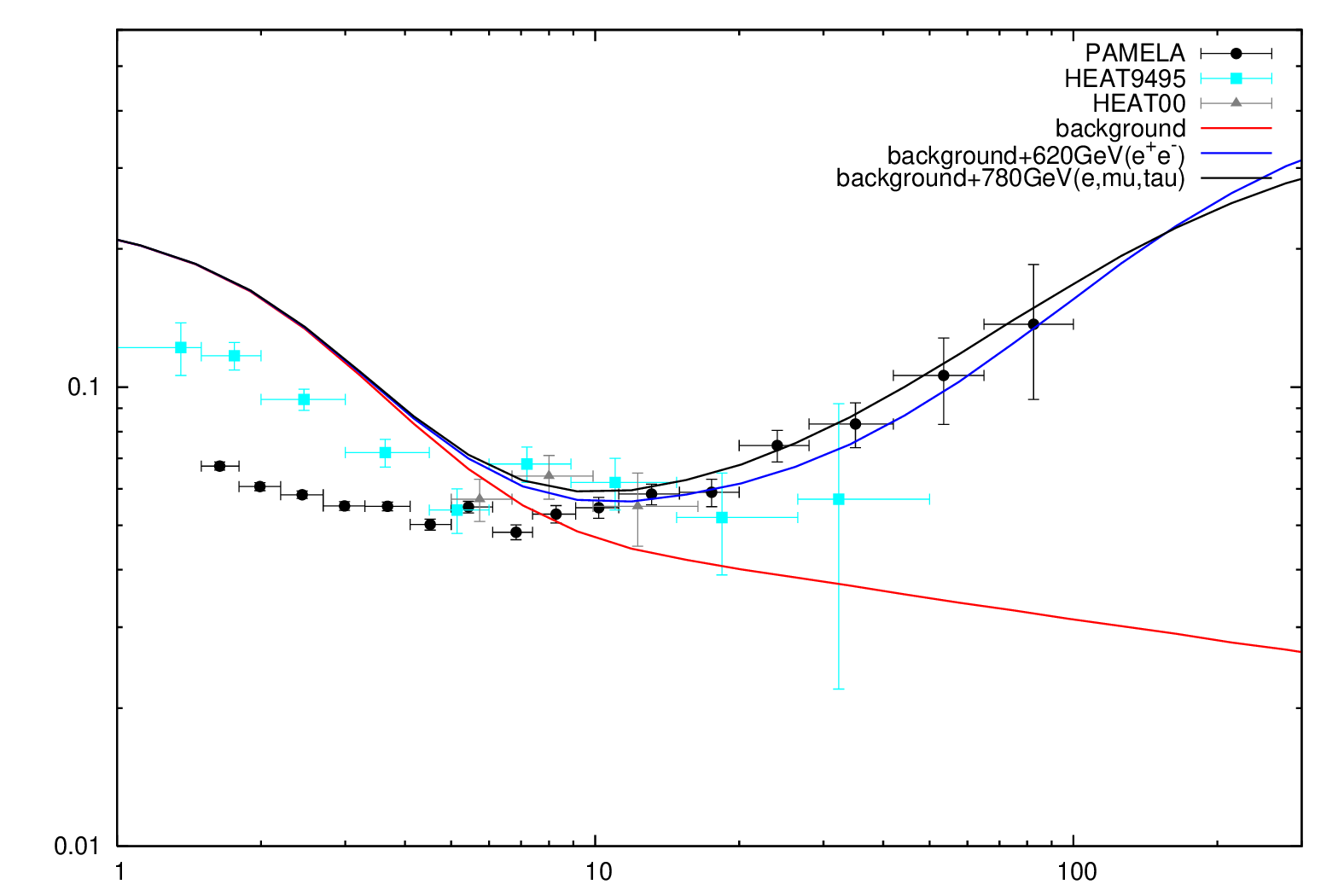}} \caption{Top:
Electron/positron spectrum including contribution from DM
annihilation compared with the ATIC/PPB-BETS data; Bottom:
$e^+/(e^-+e^+)$ including contribution from DM annihilation as
function of energy compared with the PAMELA and HEAT data. Two
models are considered: in one model DM mass is $620$ GeV with
$e^+e^-$ being the main annihilation channel, while in the other
model DM mass is $780$ GeV with equal branching ratio into
$e^+e^-$, $\mu^+\mu^-$ and $\tau^+\tau^-$. } \label{data}
\end{center}
\end{figure}

In Fig. \ref{data} we show our results compared with the ATIC/PPB-BETS
and PAMELA/HEAT
data. Two models are adopted to account for the data. In one model
we assume the dark matter mass $620$ GeV and dark
matter annihilates into electron/positron pairs dominantly.
In the other model we take the DM mass as $780$ GeV and assume
DM annihilates into $e^+e^-$, $\mu^+\mu^-$ and $\tau^+\tau^-$ with
 equal branching ratios.
We can see that both models can give good fit to the data. The value
$BF<\sigma v>$ is taken as $7.2\times 10^{-24}\, \textrm{cm}^3
\textrm{s}^{-1}$ and $1.8\times 10^{-23} \,\textrm{cm}^3
\textrm{s}^{-1}$ respectively for the two models, corresponding to
the boost factor $240$ and $600$ respectively if we assume DM
produced thermally with natural value $<\sigma v> = 3 \times
10^{-26} \,\textrm{cm}^3 \textrm{s}^{-1}$. In order to explain the
ATIC/PPB-BETS and PAMELA/HEAT data we consider the scenario where
the DM is produced by some nonthermal mechanism~\cite{zhang}.
According to Eqs. (\ref{cross-section-majorana-3}) and
(\ref{cross-section-dirac-3}) with the Yukawa couplings
$y_{\alpha,\beta }<\sqrt {4\pi}$, we can see that: (i) for the
Majorana fermion as dark matter, the required boost factor should be
larger than ${\cal O}(10^{7})$, which is in lack of reasonable
explanation; (ii) for the Dirac fermion as dark matter, the boost
factor can be absent or can be as small as ${\cal O}(10)$, which may
be due to the clumps of DM
distribution~\cite{yuan}. %In addition, if the Yukawa couplings
%$y_{\alpha }$, $h_{\alpha }$ are about 3 which is still smaller than
%$\sqrt {4\pi}$ for the violation of perturbation, we can naturally
%explain the ATIC/PPB-BETS and PAMELA/HEAT experiments without
%introducing any boost factors.

%%%%%%%%%%%%%%%%%%%%%%%%%%%%%%%%%%%%%%%%%%%%%%%%%%%%%%%%%%%%%%%%%%%%%%

%%%%%%%%%%%%%%%%%%%%%%%%%%%%%%%%%%%%%%%%%%%%%%%%%%%%%%%%%%%%%%%%%%%%%%

\subsection{Summary}

The recent observations by ATIC/PPB-BETS and PAMELA imply that DM
should dominantly annihilate into the leptons. This phenomenon
inspires us to connect the leptonic annihilation of the DM with
other leptonic processes, such as the neutrino mass-generation. In
some radiative seesaw models, where the neutrinos obtain their
small Majorana or Dirac masses at one-loop order, DM can naturally
have dominant channel annihilating into the leptons. We show that
the Dirac radiative seesaw model can account for the observed
electron/positron excesses even if we don't introduce large boost
factor. However, the Majorana case is strongly disfavored due to a
unreasonably large boost factor to account for the PAMELA/ATIC
data. With small modifications, we might explain the INTEGRAL
experiment via the exciting DM. Furthermore, we constructed one
kind of models with traditional seesaw mechanism. These models can
not only explain the ATIC/PPB-BETS and PAMELA cosmic-ray
experiments similarly, but also explain the INTEGRAL experiment.

\acknowledgements This work was supported in part by the Natural
Sciences Foundation of China (Nos. 10575111, 10773011, 10821504, 90303004, 10533010, 10675136),
by the Chinese Academy of Sciences under the grant No.
KJCX3-SYW-N2, and by the Cambridge-Mitchell Collaboration in
Theoretical Cosmology.


\begin{thebibliography}{99}


\bibitem{Chang:2008zz}
J. Chang {\it et al.}, Nature {\bf 456}, 362 (2008).


\bibitem{Torii:2008}
S. Torii, {\it et al.}, arXiv:0809.0760 [astro-ph].

\bibitem{Adriani:2008zr}
O. Adriani {\it et al.}, arXiv:0810.4995 [astro-ph].


\bibitem{Adriani:2008zq}
O. Adriani {\it et al.}, arXiv:0810.4994 [astro-ph].



\bibitem{Barwick:1997ig}
S.W. Barwick {\it et al.} [HEAT Collaboration], Astrophys. J. {\bf
482}, L191 (1997). %[arXiv:astro-ph/9703192].


\bibitem{Aguilar:2007yf}
M. Aguilar {\it et al.}  [AMS-01 Collaboration], Phys. Lett.  B
{\bf 646}, 145 (2007). %[arXiv:astro-ph/0703154].

\bibitem{yin}
M.~Cirelli, M.~Kadastik, M.~Raidal and A.~Strumia,
  arXiv:0809.2409 [hep-ph];
  N.~Arkani-Hamed, D.~P.~Finkbeiner, T.~Slatyer and N.~Weiner,
  arXiv:0810.0713 [hep-ph];
  I.~Cholis, D.~P.~Finkbeiner, L.~Goodenough and N.~Weiner,
  arXiv:0810.5344 [astro-ph].
P.F. Yin, Q. Yuan, J. Liu, J. Zhang, X.J. Bi, S.H. Zhu, X.M.
Zhang, arXiv:0811.0176;
%\cite{Baek:2008nz}
\bibitem{Baek:2008nz}
 S.~Baek and P.~Ko,
 %``Phenomenology of $U(1)_{L_\mu - L_\tau}$ charged dark matter at PAMELA and
 %colliders,''
 arXiv:0811.1646 [hep-ph];
 %%CITATION = ARXIV:0811.1646;%%
 K.~Ishiwata, S.~Matsumoto and T.~Moroi,
  arXiv:0811.0250 [hep-ph];
  P.~J.~Fox and E.~Poppitz,
  arXiv:0811.0399 [hep-ph];
  C.~R.~Chen, F.~Takahashi and T.~T.~Yanagida,
  arXiv:0811.0477 [hep-ph];
  K.~Hamaguchi, E.~Nakamura, S.~Shirai and T.~T.~Yanagida,
  arXiv:0811.0737 [hep-ph];
    A.~Ibarra and D.~Tran,
  arXiv:0811.1555 [hep-ph];
J. Zhang , X.J. Bi, J. Liu, S.M. Liu, P.F. Yin, Q. Yuan, S.H. Zhu,
arXiv:0812.0522.

\bibitem{donato}
F. Donato, D. Maurin, P. Brun, T. Delahaye, P. Salati, e-Print:
arXiv:0810.5292 [astro-ph].

\bibitem{minkowski1977}
P. Minkowski, Phys. Lett. \textbf{67B}, 421 (1977); T. Yanagida, in
{\it Proc. of the Workshop on Unified Theory and the Baryon Number
of the Universe}, ed. O. Sawada and A. Sugamoto (KEK, Tsukuba,
1979), p. 95; M. Gell-Mann, P. Ramond, and R. Slansky, in {\it
Supergravity}, ed. F. van Nieuwenhuizen and D. Freedman (North
Holland, Amsterdam, 1979), p. 315; S.L. Glashow, in {\it Quarks and
Leptons}, ed. M. L$\rm\acute{e}$vy {\it et al.} (Plenum, New York,
1980), p. 707; R.N. Mohapatra and G. Senjanovi$\rm\acute{c}$, Phys.
Rev. Lett. \textbf{44}, 912 (1980).

\bibitem{mw1980}
M. Magg and C. Wetterich, Phys. Lett. B \textbf{94}, 61 (1980); J.
Schechter and J.W.F. Valle, Phys. Rev. D \textbf{22}, 2227 (1980);
T.P. Cheng and L.F. Li, Phys. Rev. D \textbf{22}, 2860 (1980); G.
Lazarides, Q. Shafi, and C. Wetterich, Nucl. Phys. B \textbf{181},
287 (1981); R.N. Mohapatra and G. Senjanovi$\rm\acute{c}$, Phys.
Rev. D \textbf{23}, 165 (1981).



\bibitem{ma1998}
E. Ma, Phys. Rev. Lett. \textbf{81}, 1171 (1998); and references
therein.


\bibitem{ma2006}
E. Ma, Phys. Rev. D \textrm{73}, 077301 (2006); J. Kubo, E. Ma, D.
Suematsu, Phys. Lett. B \textbf{642}, 18 (2006).

\bibitem{gs2007}
P.H. Gu and U. Sarkar, Phys. Rev. D \textbf{77}, 105031 (2008).

\bibitem{gs2008}
P.H. Gu and U. Sarkar, Phys. Rev. D \textbf{78}, 073012 (2008).


%%%%%%%%%%%%%%%%%%%%%%%%%%%%%%%%%%%%%%%%%%%%%%%%%%%%%%%%%%%%%%%%%%%%%%%%%%%%%%

%%%%%%%%%%%%%%%%%%%%%%%%%%%%%%%%%%%%%%%%%%%%%%%%%%%%%%%%%%%%%%%%%%%%%%%%%%%%%%

%\cite{Churazov:2004as}
\bibitem{Churazov:2004as}
  E.~Churazov, R.~Sunyaev, S.~Sazonov, M.~Revnivtsev and D.~Varshalovich,
  %``Positron annihilation spectrum from the Galactic Center region observed by
  %SPI/INTEGRAL,''
  Mon.\ Not.\ Roy.\ Astron.\ Soc.\  {\bf 357}, 1377 (2005);
%  [arXiv:astro-ph/0411351].
  %%CITATION = MNRAA,357,1377;%%
G.~Weidenspointner {\it et al.},
  %``The sky distribution of positronium annihilation continuum emission
  %measured with SPI/INTEGRAL,''
  arXiv:astro-ph/0601673;
  %%CITATION = ASTRO-PH/0601673;%%
G.~Weidenspointner {\it et al.},
  %``The sky distribution of 511 keV positron annihilation line emission as
  %measured with INTEGRAL/SPI,''
  arXiv:astro-ph/0702621.
  %%CITATION = ASTRO-PH/0702621;%%




%\cite{Finkbeiner:2007kk}
\bibitem{Finkbeiner:2007kk}
  D.~P.~Finkbeiner and N.~Weiner,
  %``Exciting Dark Matter and the INTEGRAL/SPI 511 keV signal,''
  Phys.\ Rev.\  D {\bf 76}, 083519 (2007).
%  [arXiv:astro-ph/0702587].
  %%CITATION = PHRVA,D76,083519;%%



%\cite{TuckerSmith:2001hy}
%\bibitem{TuckerSmith:2001hy}
%  D.~Tucker-Smith and N.~Weiner,
  %``Inelastic dark matter,''
%  Phys.\ Rev.\  D {\bf 64}, 043502 (2001);
%  [arXiv:hep-ph/0101138].
  %%CITATION = PHRVA,D64,043502;%%
%``The status of inelastic dark matter,''
%  Phys.\ Rev.\  D {\bf 72}, 063509 (2005);
%  [arXiv:hep-ph/0402065].
  %%CITATION = PHRVA,D72,063509;%%
%S.~Chang, G.~D.~Kribs, D.~Tucker-Smith and N.~Weiner,
  %``Inelastic Dark Matter in Light of DAMA/LIBRA,''
%  arXiv:0807.2250 [hep-ph].
  %%CITATION = ARXIV:0807.2250;%%


%\cite{Bae:2008sp}
\bibitem{Bae:2008sp}
  K.~J.~Bae, J.~H.~Huh, J.~E.~Kim, B.~Kyae and R.~D.~Viollier,
  %``White dwarf axions, PAMELA data, and flipped-SU(5),''
  arXiv:0812.3511 [hep-ph].
  %%CITATION = ARXIV:0812.3511;%%


%\cite{Jiang:2008yf}
\bibitem{Jiang:2008yf}
  J.~Jiang, T.~Li, D.~V.~Nanopoulos and D.~Xie,
  %``F-SU(5),''
  arXiv:0811.2807 [hep-th].
  %%CITATION = ARXIV:0811.2807;%%


%%%%%%%%%%%%%%%%%%%%%%%%%%%%%%%%%%%%%%%%%%%%%%%%%%%%%%%%%%%%%%%%%%%%%%%%%%%%%%

%%%%%%%%%%%%%%%%%%%%%%%%%%%%%%%%%%%%%%%%%%%%%%%%%%%%%%%%%%%%%%%%%%%%%%%%%%%%%%


\bibitem{Gaisser:1990vg}
T.K. Gaisser, {\it Cambridge, UK: Univ. Pr. (1990) 279 p}


\bibitem{Strong:1998pw}
A.W. Strong and I. V. Moskalenko, Astrophys. J. {\bf 509}, 212
(1998). [arXiv:astro-ph/9807150].


\bibitem{galprop}
http://galprop.stanford.edu/

\bibitem{nfw}
J.F. Navarro, C.S. Frenk, S.D.M. White, Astrophys. J. {\bf 490}, 493
(1997).

\bibitem{zhang}
W.B. Lin, D.H. Huang, X. Zhang, R.H. Brandenberger,
Phys. Rev. Lett. \textbf{86}, 954 (2001);
R. Jeannerot, X. Zhang and R. Brandenberger, JHEP 9912, 003 (1999).

\bibitem{yuan}
Q. Yuan, X.J. Bi,
JCAP 0705, 001 (2007);
J. Lavalle, Q. Yuan, D. Maurin, X.-J. Bi
A\&A \textbf{479}, 427 (2008).




\end{thebibliography}
\end{document}